\documentclass[12pt,preprint]{aastex}

\shorttitle{Small scale systems of galaxies I}
\shortauthors{Tanvuia et al.}

\begin{document}


\title{Small scale systems of galaxies. I. \\ 
       Photometric and spectroscopic properties of 
       members\footnote{Based on observations obtained at the 
       European Southern Observatory, La Silla, Chile (Programme Nr. 57.B--036)}}


\author{L.~Tanvuia}
\affil{Institut f\"ur Astronomie, Universit\"at Wien,
       T\"urkenschanzstra{\ss}e 17, A-1180 Wien, Austria}
\email{tanvuia@astro.univie.ac.at}

\author{B.~Kelm, P.~Focardi}
\affil{Dipartimento di Astronomia, Universit\`a di Bologna, 
       Via Berti Pichat 6, Bologna, Italy}
\email{kelm@bo.astro.it, focardi@bo.astro.it}

\author{R.~Rampazzo}
\affil{INAF - Osservatorio Astronomico di Padova, Vicolo dell'Osservatorio 5, 
       I-35122, Padova, Italy}
\email{rampazzo@pd.astro.it}

\and

\author{W.W.~Zeilinger}
\affil{Institut f\"ur Astronomie, Universit\"at Wien,
       T\"urkenschanzstra{\ss}e 17, A-1180 Wien, Austria}
\email{zeilinger@astro.univie.ac.at}


\begin{abstract}

This paper is the first of a series addressed to the investigation of
galaxy formation/evolution in small scale systems of galaxies (SSSGs)
which are located in low density cosmic environments.  Our algorithm
for SSSG selection, includes galaxy systems of 2 or more galaxies
lying within $\Delta$$cz$$\leq$ 1000 km~s$^{-1}$ and a 200 $h_{100}^{-1}$ kpc
radius volume.  We present the analysis of the photometric and
spectroscopic properties of 19 member galaxies belonging to a sample
of 11 SSSGs.

In the $\mu_e$ -- $r_e$ plane, early--type members may be considered
``ordinary'', not ``bright'' galaxies in the definition given by
\citet{cap92} with a significant fraction of galaxies having a disk or
disky isophotes. We do not detect fine structure and signatures
of recent interaction events in the early--type galaxy population, a
picture also confirmed by the spectroscopy.

At odd, there are several spiral members with open arm configurations
as expected in interacting systems. At the same time, emission lines
in the spectra of spiral members fall in the HII regions regime defined
with diagnostic diagrams \citep{vei87}.  None of the objects
displays unambiguous indication of nuclear activity, although four
spiral nuclei could be ascribed to the class of Seyferts. The star
formation rate seems enhanced over the average expected in spiral
galaxies only for poorer SSSGs in particular pairs ($\leq$50
M$_\odot$~yr$^{-1}$) but without being in the range of starburst systems.

\end{abstract}




\section{Introduction}
Among cosmic environments, small scale systems of galaxies (SSSGs) in
the field are those in which not only galaxy--galaxy interactions but
also the evolution of cosmic structures can be studied at a
``cellular'' level. To the SSSG class of cosmic structure we may
ascribe galaxy systems with different richness and density
characteristics spanning from compact groups to poorer configurations
like galaxy pairs.

Together with compact galaxy groups \citep[and references
therein]{hic97}, investigated in definitely more detail, recent X-ray
observations suggest the physical reality of some loose groups
\citep{pon96, mul00} and even pairs \citep{hen99, tri01}.  The X-ray
diffuse component and the plethora of faint galaxies associated to the
few dominant members is interpreted as due to the presence of a deep
potential well \citep{mul00}.

Are then SSSGs long lasting associations or, in the hierarchical
evolution scenario, the debris of older, richer and sparse
configurations?  Are pairs the debris of a pristine group and
consequently a way station toward isolated Es? NGC 1132, an isolated
elliptical with extended X-ray diffuse emission \citep{mul99}, could
be a prototypical example of the evolution of such systems. Some
scenarios \citep{dia94, gov96} depict rich and compact galaxy
structures, like Hickson compact groups, as substructures of larger
ones, with different degree of equilibrium.  \citet{zab98} found
typically 20 to 50 dwarf galaxies associated to their X-ray detected
groups. In this context, SSSGs of different richness and degree of
compactness have to be studied comparatively as ``single'' class of
cosmic environments.

The understanding of galaxy--galaxy interaction phases and the
evolution of cosmic structures are deeply interconnected. Galaxy
encounters in SSSGs are less frequent than in galaxy clusters, but the
low velocity dispersion of the SSSG may lead to efficient merging
episodes \citep{bar96}. Interactions could severely alter the
properties of a galaxy up to modify the original morphological class
\citep{bar96, ken96} and then re-direct the evolution of a galaxy
triggering various phenomena ranging from star formation \citep{lon99}
to galaxy activity \citep{mon94, raf95, lau95, kee96, kel98,
coz00}. In this picture, the environment plays a key role since it
dictates the predominant type of encounters \citep{moo96, bar96}. A
connection should then exist between the local environment and the
global properties of galaxies inhabiting it.  For rich galaxy systems,
many aspects of the connection between environment and galaxy
properties have emerged, from radio \citep{hay84} to X-ray properties
\citep{for82}.  The link between the environment and the galaxy
evolution in SSSGs is still not fully understood \citep{pon96, lon98a,
lon98b, lon99, lon00, ram00, mul00, coz00} and partly suffers a
shortage of information with respect to richer environments.

In order to understand which parameters dominate in galaxy
interactions, we started a study of SSSGs, defined in 3D redshift
space. SSSGs have been selected in a low density environment,
i.e.~they appear as overdense systems with respect to the average
galaxy distribution.

The present paper collects photometric and spectrophotometric data for
the main members of a sample of SSSGs. Within the picture described
above, dominant galaxies in SSSG have potentially the power of
revealing the history and evolutionary phase of the respective
SSSGs. Elliptical galaxies may be at the center of the potential well
of a SSSG as suggested by X-ray observations \citep{mul00} and could
present fine structures \citep{sch96} reminiscent of past interaction
events. Spiral galaxies in mixed pairs could be the last un--digested
member of a pre--existing group \citep{ram92} and could reveal their
on--going interaction not only through morphological distortion but
also in their possibly triggered activity.

The goals of the present paper are the following: 1) to obtain a
redshift estimate which will provide an independent check of the
systemic velocities of the SSSG members typically made available by
large redshift surveys, 2) to investigate the structure of the member
galaxies through a detailed surface photometry, 3) to analyze the
possible induced activity using medium resolution spectroscopy and
diagnostic models.

The paper is organized as follows: The sample is defined in \S~2. A
description of the photometric and spectroscopic observations, data
reduction and analysis is given in \S~3, while results of individual
objects are presented in \S~4. The global morphological, photometric
and spectroscopic properties of the present sample as a function of
the SSSG and SSSG member properties are discussed in \S~5.

\section{The sample definition}

The SSSG sample includes 11 systems of 2 or more galaxies lying at
similar redshift ($\Delta$cz $\leq$ 1000 km/s) and within a 200
$h^{-1}$\,kpc radius area, for which new spectroscopic or photometric
data have been acquired.

These confirmed SSSGs constitute a small subset of a SSSGs candidate
sample which includes likely misclassified isolated galaxies in ZCAT
\citep{huc92}. The version of ZCAT used for selection of
candidates contains 57536 entries.  The SSSGs candidate sample was
selected with a two-step procedure: first, isolated galaxies were
identified in ZCAT with an automatic code defining as isolated all
those galaxies presenting no neighbour(s) with known redshift within
$\Delta$cz $\pm$ 1000 km/s and $\Delta$R = 1$h^{-1}$ Mpc.  With this
procedure, 3890 galaxies were selected in the the redshift interval
$3000 < cz < 10000$ km~s$^{-1}$ out of a sample of 18677 galaxies in
total.  ZCAT is essentially a redshift compilation rather than a
complete sample and also includes a $\approx$25\% fraction of galaxies
with unknown redshift, Therefore, many of the selected isolated
galaxies presented one or more projected companion(s), and are
therefore likely misclassified isolated galaxies.  To account for
this, the code has identified, in the second step, among the isolated
galaxies, those presenting nearby projected neighbours
(i.e. neighbours whose redshift is unknown), which are clearly likely
to represent in fact small scale galaxy systems. This resulted in 423
candidate SSSGs.

All candidates lying within 1 Abell radius and $\Delta$cz $\leq$ 1000
km/s from ACO clusters \citep{str99} have been excluded from the
sample.  We have also inspected DSS images to select among the
projected pairs and groups those appearing outside clusters and dense
large groups, and are consequently suited for photometric and spectral
inspection.  Once systems inside clusters are rejected one can safely
assume the SSSG sample includes only systems in low density
environment. However, we expect that SSSGs reside in environments not
as sparse as those typically associated to single galaxies, based on
the result by \citet{foc02}, showing (in their figure 10) that compact
groups display a significant excess of neighbours with respect to
galaxies lacking a close neighbour.

The sample of galaxy systems we present here is very small: it covers
a large range in radial velocity and absolute magnitude, and is
partially based on catalogues that are not complete.  Further it has
been defined making also use of non--automatic selection criteria,
causing an intrinsic and hardly quantifiable bias.  Therefore, we do
not expect the sample to be representative of all galaxy systems in
low density environment.  It seems however, that the sample still
provide enough information that some general conclusions might be
drawn, in particular concerning the different behaviours of
early--type and late--type galaxies in pairs and groups detected in
low density environments.

The main data for the 11 confirmed SSSGs investigated here are
presented in \ref{table-1}.  We have checked the NED for SSSGs
neighbours with known redshift out to a distance of
600$h_{100}^{-1}$kpc and 1$h_{100}^{-1}$ Mpc respectively (see Table
\ref{table-1}).  Though obtained from a data base which is clearly not
complete, this information is listed to allow a quantitative first
order approximation of the galaxy density around each SSSG. All SSSGs
are in environments including less than 10 galaxies within a projected
area of 1h$_{100}^{-1}$Mpc and a redshift range of $\pm$1000
km~s$^{-1}$ from the SSSG center. This is used to argue that SSSGs are
indeed in a low density environment, ranging from completely isolated
to that typical for loose groups.

The observed sample contains additionally 2 SSSGs which have not been
investigated in detail, namely CGCG 054-011 ($\alpha$(2000) 17 15 08.7
$\delta$(2000) 08 27 22) and CGCG 054-013 ($\alpha$(2000) 17 15 11.9
$\delta$(2000) 08 25 35). According to literature data, they satisfied
our selection criteria. However, our spectroscopic investigation
revealed that they are an optical alignment having
V$_{hel}$=10001$\pm$11 km~s$^{-1}$ and V$_{hel}$=6444$\pm$11
km~s$^{-1}$ respectively. They are no further considered in the paper.

\section{Observation and reduction}

Grey-scale images of the SSSG sample obtained with the 0.91m Dutch
telescope at ESO La Silla are presented in Figure \ref{figure-1}.

\subsection{Spectroscopic observations and analysis}

Long-slit spectra were acquired at ESO 1.52m telescope at La Silla
equipped with a Boller \& Chivens Cassegrain spectrograph during an
observing run in 1996. The spectra of the SSSG member galaxies were
obtained in the wavelength region 3500 $\leq \lambda \leq$ 11000 \AA\
with a dispersion of 3.69 \AA\ pixel$^{-1}$. The slit (slit width
2\arcsec) has been oriented along the line connecting the nuclei of
two of the SSSG members in order to optimize exposure times. The
detector used was a FA 2048L UV--coated CCD (ESO CCD \#15.)  The
members which have been observed within each SSSG are reported in
Table \ref{table-2}. The observing log of the observations is given in
Table \ref{table-3}.

The spectra were calibrated with the ESO-MIDAS\footnote{ESO-MIDAS is
developed and maintained by the European Southern Observatory.}
software package using standard procedures for bias subtraction and
flat field correction. Artifacts produced by cosmic ray events were
removed by applying a filtering algorithm. Wavelength calibration was
performed on the frames by fitting a third order polynomial, using as
reference the helium--argon spectrum taken before each object
spectrum.  The spectra were flux calibrated with the
IRAF\footnote{IRAF is distributed by the National Optical Astronomy
Observatories, which are operated by the Association of Universities
for Research in Astronomy Inc., under cooperative agreement with the
National Science Foundation.} package KPNOSLIT. This procedure
included the airmass and extinction corrections using standard stars
and the atmospheric extinction coefficients.  The flux calibrated,
de--redshifted spectra are presented in Figure \ref{figure-2}.

Heliocentric systemic velocities have been obtained through a 
cross--correlation technique using the IRAF task CROSSCOR or through the
interpolation of emission (or absorption) lines with a single gaussian
fit, when the cross--correlation yielded no results because of low
signal--to--noise absorption line data.  Our redshift measurements
were compared with data available in the literature.  We found an
average systemic velocity difference of 84 km~s$^{-1}$ with a standard
deviation of 132 km~s$^{-1}$.  

The integrated flux of the prominent emission lines (H$\beta$, 
[OIII] $\lambda$4959\AA, $\lambda$5007\AA, 
H$\alpha$, [NII] $\lambda$6548\AA, and [SII] $\lambda$6717\AA) was 
measured using an interactive gaussian fitting procedure
and is reported in the Table \ref{table-4}.

\subsection{Photometric observations and analysis}

Imaging was carried out with the 0.91m Dutch telescope at ESO, La
Silla, Chile, during a single run in 1996.  Bessel $R$ band images
were obtained under homogeneous observing conditions. The detector
employed was a SITe 512$\times$512 pixel CCD (ESO CCD \#33) with a
scale of 0.442$''$~pixel$^{-1}$ yielding a field of view of
3\arcmin.8$\times$3\arcmin.8. The image cleaning, dark and bias
subtraction, flat--fielding, cosmic ray removal and the final
calibration and image manipulation were performed using the ESO-MIDAS
software package.

Standard stars used for calibration purposes were obtained in  the
same instrumental set--up  in the fields of PG1633+099 (4 stars),
SA 110 (4 stars), Markarian~A (4 stars),  T--Phe (3 stars). Each of
these fields contains several calibration stars, avoiding the crowding
problem  typical of fields surrounding globular cluster areas. 
We used the standard photometric transformation equations tailored
to the ESO (La Silla) system of extinction coefficients. 

Ellipses were fitted by weighted least squares to the isophotes of the
SSSG member galaxies using the IRAF ISOPHOTE package \citep{jed87} 
within STSDAS. Surface brightness, position angle and ellipticity
profiles together with the Fourier--coefficients to quantify the
deviations of the isophotes from pure ellipses were derived for the
early--type members. The $b_4$ coefficient is in particular used to
analyze the boxiness/diskiness of the early-type galaxies.  In the
case of late--type disk galaxies only the surface brightness,
ellipticity and position angle profiles are used for the further
analysis. Furthermore, in order to avoid artifacts due to the presence
of spiral arms and bright HII regions, the ellipse fitting procedure
was repeated iteratively until a satisfactory representation of the
stellar disk component could be achieved. Using the parameters
extracted from the surface photometry, a smooth model of the galaxy
has been produced and subtracted from the original image in order to
evidence faint structures \citep{sch92} possibly connected to
on--going/past interaction episodes. The profiles outside the seeing
dominated central region are presented in Figures \ref{figure-3} and
\ref{figure-4}. The observing log and the results from surface
photometry are summarized in Table \ref{table-5}. Further, the imaging
data was used to establish morphological types for those galaxies,
where either no classification is available in NED or the present data
reveal a a morphology in conflict with NED.

\section{Results and Comments on individual objects}

In this section we comment about particular features of the SSSGs in
the sample and on individual galaxies in the SSSGs recovered from
surface photometry and spectroscopy.

\noindent\underbar {\bf SSSG~1} 
The galaxies are members of WBL~637 \citep{whi99} which is formed by 4
objects of which SSSG1a and SSSG1b are WBL members 002 and 003,
respectively. The WBL determination agrees with our redshift study of
possible neighbours as shown in Table 1. Both galaxies have
early--type morphology. No emission lines are detected in the spectra
of both members.  Surface photometry of SSSG~1a reveals an
early--type disk system seen nearly edge--on. The stellar disk
component appears prominent both in the residual image and the $b_{4}$
profile. SSSG~1b is an elliptical galaxy, which shows a moderate,
$\approx$ 10$^\circ$, isophote twist, while the ellipticity increases
up to $\approx$0.4. The Fourier coefficient $b_{4}$ reveals a small
disk component in the range 4\arcsec $\leq r \leq$ 10\arcsec. The
presence of the inner stellar disk is also visible in the surface
brightness profile. A diffuse structure is present in the residual
image.  A plethora of dwarf galaxies is detected nearby
SSSG~1b. According to the systemic velocities of SSSG~1a and SSSG~1b,
we suggest that the objects are physically paired, although we do not
notice obvious signatures of interaction. Spectra, shown in Figure
\ref{figure-2}, are typical of their morphological class.

\noindent\underbar {\bf SSSG~2 } 
This system is a chain of 4 bright galaxies with a systemic velocity
difference less than 140 km~s$^{-1}$. Galaxies in this poor
association appear also in the WBL catalog (WBL 642). The chain, from
the north to the south, is composed of a Sc face-on spiral with
multiple arms (SSSG~2d), followed by two early--type galaxies (SSSG~2b,
SSSG~2a). The fourth member (SSSG~2c) is a face--on Sc spiral with two
prominent open arms but diffuse patchy areas, some of them marking
incipient spiral arms. 

SSSG~2a (E/S0) shows a rising ellipticity profile and a positive (up
to 4\%) Fourier coefficient $b_{4}$ indicating the presence of an
inner stellar disk. No emission lines are detected. Three faint
objects (Figure \ref{figure-1}b) are nearby to SSSG~2a in
projection, two north and one south of it. The northern object of the
two is compact, with faint and narrow emission lines. The
H$\alpha$/[NII] $\lambda$6583\AA\ line ratio indicates a Seyfert
1--like AGN. The radial velocity is 2159$\pm$54 km~s$^{-1}$. The
second object is blue, with faint and narrow lines. The radial
velocity is 2165$\pm$262 km~s$^{-1}$. The southern object is an
edge--on disk galaxy with a radial velocity of 6717$\pm$59 km~s$^{-1}$
suggesting a physical association with SSSG~2a. SSSG~2b (E/S0) shows a
low ellipticity $\epsilon \approx$ 0.2 and a strong isophotal twist of
$\approx$ 120$^\circ$.  The Fourier coefficients $a_{4}$ and $b_{4}$
indicate the presence of a stellar disk component in the inner part of
the galaxy.  SSSG~2c is a face--on late--type spiral galaxy with
multiple arms of which two dominate. Surface brightness and position
angle profiles reveal the presence of a small exponential bar
component in the region 4$'' \leq r \leq $ 13$''$. The residual image
shows a number of smaller spiral arms. SSSG~2d is a late--type
face--on spiral with flocculent arms. Some of them mark incipient
spiral arms.  A number of HII regions appears in the residual image
after subtracting a smooth galaxy model. SSSG~2c and SSSG~2d show
emission lines ratios characteristic of HII regions.

A number of faint galaxies is noted in the immediate surroundings of
the bright galaxies.  Of particular interest is a faint disk galaxy
(Figure. \ref{figure-1}c) which is detected between SSSG~2b
and SSSG~2d. The object shows strong and narrow emission features in
the spectrum, with line ratios consistent with typical HII regions.
The measured radial velocity is 6626$\pm$64 km~s$^{-1}$ in agreement
with the systemic velocity of SSSG~2.

\noindent\underbar {\bf SSSG~3} 
The isolated multiplet consists of 4 confirmed galaxies with 5
possible neighbours within 1 Mpc. The two early--type member galaxies
of the multiplet are seen nearly edge--on.  SSSG~3a has been
classified as E6 according to NED. However, the surface brightness
profile reveals a bulge and a stellar disk component suggesting
therefore an E/S0 type.  The presence of a stellar disk component is
further confirmed by an overall positive $b_{4}$ coefficient and
evidenciated in the residual image.  SSSG~3b has a constant position
angle profile and the ellipticity increases up to 0.6. The surface
brightness profile also shows the presence of a bulge and a disk
component supporting the classification of S0:Sp from NED. The latter
is also revealed in the residual image. The galaxy is surrounded by
faint objects possibly dwarf galaxies. No emission lines are detected
in SSSG~3a and SSSG~3b as expected from their morphological
class. There are no morphological signatures of interactions detected
between the galaxies.

\noindent\underbar {\bf SSSG~4}. 
The pair also known as KPG~551 \citep{kar72} shows pronounced
signatures of interaction. The southern object SSSG~4a appears
strongly interacting and is of irregular type.  The northern object
SSSG~4b, which is classified as SAB pec, shows an arc--like tail/arm
extending to the northeast of the galaxy.  This structure contains
numerous HII regions \citep{jun98}.  We measured a virtually null
systemic velocity difference as reported in the literature. Both
members show prominent emission lines. The apparent peculiarities of
this isolated pair prevent us to perform a detailed surface
photometry. Between the SSSG~4a and SSSG~4b there is a faint object,
with narrow emission lines. No absorption lines were detected.  The
measured radial velocity is 6954$\pm$127 km~s$^{-1}$.

\noindent\underbar {\bf SSSG~5} 
The isolated pair ($\Delta V$=104 km~s$^{-1}$) is composed of an
edge-on lenticular and a grand design spiral galaxy. The S0 galaxy
SSSG~5a shows a two component surface brightness profile consisting of
a bulge and stellar disk component. The ellipticity profile rises up
to 0.5 and the position angle profile is nearly constant at $\approx$
80$^\circ$.  The Fourier coefficient $b_{4}$ further evidences the
disk component ($\approx$ 8\%). No emission lines are detected in the
spectrum. The presence of strong spiral arms in SSSG~5b prevented a
detailed surface photometric analysis.  The galaxy spectrum shows
faint emission lines.

\noindent\underbar {\bf SSSG~6} 
The members of the multiplet appear to be quite separated from each
other.  SSSG~6 is composed of three late--type members: SSSG~6a, seen
almost face--on, SSSG~6b and a third object, for which no spectroscopic
data are available and which is seen almost edge--on.  The derived
H$\alpha$/[NII] $\lambda$6583\AA\ line ratio for the nucleus of
SSSG~6a suggests a Seyfert 1--like AGN. This galaxy is also detected
by IRAS (IRAS~00000-0359). The presence of strong spiral arms in
SSSG~6a prevent us to perform a detailed surface photometry. A number
of HII regions appears in the residual image after substraction of a
smooth model galaxy. SSSG~6b appears to be distorted with two inner
main arms and outer arms completely decoupled. The difference of their
systemic velocities is $\approx$200 km~s$^{-1}$ suggesting 
a physical association.

\noindent\underbar {\bf SSSG~7} 
The pair is composed of an edge--on lenticular with a thick disk and
an early spiral galaxy with very thin spiral arms showing patchy HII
regions. The systemic velocity difference between the members is only
$\approx$70 km~s$^{-1}$.  SSSG~7a shows a rising ellipticity and a
constant position angle profile. The $b_{4}$ profile indicates a
strong (8\%) disk component. The photometric data are influenced by a
bright star in the vicinity. No spiral arm structure is visible
suggesting therefore an S0 type instead of the classification as S? in
the NED.  SSSG~7b shows faint spiral arms after substraction of a
smooth model galaxy. These arms are more reminiscent of a sort of
shell structure created by weak interaction \citep{tho91, wei93}
rather than typical spiral arms. The H$\alpha$ intensity is
approximately similar to that of [NII] in the galaxy center which
indicates the presence of a strong H$\alpha$ component in
absorption. The H$\alpha$/[NII] $\lambda$6583\AA\ line ratio indicates
an AGN of Seyfert~1 type.

\noindent\underbar {\bf SSSG~8} 
The loose pair consists of two barred spirals seen almost face-on. The
imaging data reveal a population of faint objects, possibly dwarf
galaxies, surrounding the bright SSSG members. A spectrum was
obtained only for the SSSG~8b, which reveals emission lines typical
for HII regions.

\noindent\underbar {\bf SSSG~9} 
The triplet is composed of an unperturbed elliptical and two spiral
galaxies.  SSSG~9a is a face--on spiral  showing multiple arm
structure, a small bulge and a prominent disk component. The analysis
of the surface brightness profile reveals the presence of an elongated
component (possibly a bar component) in the range 10\arcsec $\leq r \leq$
20\arcsec.  The spiral arms are very prominent with a number of large
HII regions particulary in the southern arm.  SSSG~9b is an apparently
undisturbed E0.  A spectrum was obtained of the elliptical (SSSG~9b)
and the brighter spiral galaxy (SSSG~9a). The difference of the
systemic velocities between these two galaxies is 122 km~s$^{-1}$. It
is noticeable that the stellar and gas components in SSSG~9a differ in
the systemic velocity of about 80 km~s$^{-1}$. The systemic velocity
of the gas component is closer to the systemic velocity of SSSG~9b.

\noindent\underbar {\bf SSSG~10}
\citet{pie00} performed a photometric study of the pair which appears
completely damaged by the encounter. They report that a complex system
of debris extend from the galaxy in south--east direction with intense
knots.  SSSG~10a is of irregular shape shows strong and narrow
emission lines in the spectrum.  SSSG~10b is a typical S0, with a dust
lane in the center and appears slightly damaged by the encounter in
the eastern outer part. The line ratio (H$\alpha$/[NII]
$\lambda$6583\AA)$<$1 of the SSSG~10b is consistent with that of
typical Seyfert galaxy, no [OIII] $\lambda$5007{\AA} emission is detected.

\noindent\underbar {\bf SSSG~11} 
The pair does not present evident signatures of interaction as also
reported by \citet{pie00}. SSSG~11a is a late--type spiral seen
edge--on.  Strong emission lines are detected in the spectrum. Also
absorption lines are present, but they appear very faint. The northern
member, SSSG~11b, is a faint spiral. The spectral features are too
faint to be further analyzed.

\section{Properties of the morphological classes}

\subsection{Properties of early--type galaxies}

The early--type galaxies of the SSSG sample have been searched for the
presence of fine structures, using the scheme developed by
\citet{sch92}. However, no significant structures could be
detected. This may indicate that either these galaxies remained
undisturbed during recent interactions within the SSSG or the
interactions occured only on small scale. Such events could be mass
accretions of (gas--rich) dwarf galaxies resulting in kinematically
decoupled components which are frequently observed in early--type
galaxies. The presently available observational material however has not
sufficient resolution to answer this question.

The Hamabe--Kormendy relation HK87 \citep{ham87} is a useful tool to
study distributions of early--type galaxies using the classification
by \citet{cap92}, hereafter CCD92, in ordinary and bright
classes. Bright galaxies are defined as those having M$_B$ $<$ -19.3
and R$_e$ $>$ 3 kpc (H$_0$=70 km~s$^{-1}$~Mpc$^{-1}$) while
$\mu_e$=2.94log R$_e$ + 20.75 traces the HK87 relation.  CCD92 show
that ordinary galaxies do not tend to distribute along the HK87
relation but fill the plane for R$_e$ $<$ 3 kpc. The ordinary galaxy
class is considered as a sort of ``genetic variety'' since it
contributes to generate bright galaxies through merging processes
\citep{nav90}. In order to transform our data from R band to B band we
adopt a (B--R) = 1.5, i.e. the color of a SSP of solar metallicity
(Z=0.02) and of an age of 13--15 Gyr.  Figure \ref{figure-5} shows the
$\mu_e$ -- log R$_e$ plane. The vast majority of our objects resides
in the range of ordinary galaxies. This is not the general case of
galaxies in low density environments as shown by \citet{ram00}. They
noticed also that there is a number of bright galaxies well below the
relation.  These latter have been interpreted as a transient phase
since they include strongly interacting members. These galaxies
completely lack in our sample. This may suggest that SSSGs are either
accordant redshift unrelated galaxies or that they are still in a
stage of pre-coalescence and have not yet produced the bright merger
remnants.

\subsection{Properties of spiral galaxies}

The late--type galaxies of the SSSG sample show clear signatures of
on--going interaction in contrast to the early--types.  Following the
classification of \citet{elm87} (see Table \ref{table-6} for
classification) most spiral galaxies belong to the grand design class
typically found in the densest group environments. At the same time, a
significant number of disk galaxies is found in comparatively
low--density environments as suggested by the rather small percentage
of barred galaxies present in the sample if compared with those
detected in binary samples by \citet{elm82} and \citet{red95}.  A
large fraction of spirals, which are not seen edge--on, show open arms
indicating an on-going interaction. According to \citet{nog86} models,
open arms develop in the very early phases of an encounter. Star
formation rate reaches the maximum value of about 8 times as large as
the pre--encounter value at about $3 \times 10^8$ years after the
perigalactic passage of the perturber. In this context we suggest that
our mutiplet SSSGs could be early phases in the hierarchical
assembling process.

The emission line spectra were used to study the ionisation
mechanisms.  The emission line intensity ratios, as defined by
\citet{vei87}, [OIII]/H$\beta$ vs. [NII]/H$\alpha$ and [OIII]/H$\beta$
vs.  [SII]/H$\alpha$ of the sample galaxies are plotted in Figure
\ref{figure-6}.  \citet{vei87} defined the intrinsic flux ratio for
HII region--like objects I(H$\alpha$)/I(H$\beta$) = 2.85 and adopt for
the intrinsic ratio of AGNs I(H$\alpha$)/I(H$\beta$) = 3.1. The values
of the flux ratio in the sample are found to be generally bigger and
using the two definitions of \citet{vei87}, the value of the flux
intensity of H$\beta$ is corrected. This is attributed to the fact
that the H$\beta$ flux is underestimated due to the presence of an
absorption line component. The new value of the flux intensity of
H$\beta$ is used to compute new [OIII]/H$\beta$ vs. [NII]/H$\alpha$
and [OIII]/H$\beta$ vs. [SII]/H$\alpha$ ratios (see errorbars in 
Figure \ref{figure-6}). Figure \ref{figure-6} shows that late--type
galaxies in the sample are dominated by photoionisation: all objects
lie in the region of the diagram, which is dominated by
HII--regions. This is the case also for morphologically peculiar
galaxies, as e.g. SSSG~4b and SSSG~6b and even strongly distorted
galaxies, e.g. SSSG~4a and SSSG10a.

Star formation rates (SFR) were calculated, using the calibration
between SFR and H$\alpha$ fluxes adopted by \citet{ken98} (see Table
\ref{table-5}):

$SFR (M_\odot yr^{-1})= 7.9 \times 10^{-42} \times L_{H_\alpha} (erg s^{-1})$

Table \ref{table-4} (column 12) collects the SFR for the disk galaxies
of the sample.  Most disk galaxies (9 out of 14) can be considered
``normal'', with star formation rates up to 8 M$_\odot$~yr$^{-1}$.  The
remaining 5 objects have a significantly higher star formation rate.
However only one exceeds the value of 50 M$_\odot$~yr$^{-1}$, which
indicate a major starburst phenomenon.  It is worth remarking that the
5 disk galaxies presenting the higher SFR are hosted in the more
isolated SSSGs of the sample (see Table \ref{table-1}), and in SSSGs
containing only spiral galaxies.

\section{Summary}

In this paper we have studied 19 members in 11 SSSGs and presented
their photometric and spectroscopic properties.  Our selection
criteria have included small systems ranking from isolated pairs (KPG
551, RR23 and RR45) with no neighbours of similar luminosity to poor
groups in low density environment.  The groups are high density
configurations and some of them are catalogued in the literature as
nearby poor clusters, as in the case of SSSG~1 (WBL 637) and SSSG~2
(WBL 642).

Although the relatively small number of objects studied so far does
not allow to draw statistically significant conclusions some trends
can still be pointed out.  Early--type galaxies in SSSGs, even those
which are found in more compact multiplets, do not show relevant
signatures of interaction.  The lack of evidence of early--type
galaxies being the result of major merger events is a feature in
common with HCGs, which have a low fraction of merging candidates and
no evidence of enhanced far--infrared emission \citep{zep91, zep93,
ver98}. Further, the early--type galaxies of our sample do not show
fine structure.  This is not surprising, indeed \citet{red96} found
that e.g.~shell structures are found four times less frequently in
interacting pairs than in isolated objects of the same class. This may
be attributed to two causes. First, none of our early--type members is
suffering or has recently suffered a strong interaction episode. Most
of our early--type members have a disk component or are disky according
to the shape parameter $b_4$ which is often larger than 2\%.  Second,
the galaxies suffer ``weak'' but multiple interactions that are likely
to destroy shell structures \citep{tho90}.

At variance with early--type galaxies spirals displaying patterns
typical of ongoing interaction and high star formation rate are found
in SSGSs.  These SSGSs are the most isolated in the sample and
additionally are the only ones displaying a total spiral population.
This finding confirms previous results from studies of classical pairs
in the \citet{kar72} catalog \citep{xus91} and pairs catalogs in the
southern hemisphere \citep{com94} reporting significant enhancements
of star formation due to the infall of fresh gas triggered by the
interaction.  At the same time the lack of evidence for enhanced star
formation in spirals in the densest SSSGs agrees with previous results
on HCGs \citep{sul93,ver98} and on UZC--CGs \citep{kel03}.  Neither our
SSSGs which are strongly interacting nor members in multiplets do show
unambiguously the presence of nuclear activity.

Further steps in the study of our sample of SSSGs are
planned/on--going. Among these, to obtain X-ray imaging and a deeper
mapping of SSSGs through wide field imaging data which will permit a
better definition of their environment. The spectroscopic information
are then a necessary information to understand the significance of the
population of faint galaxies accompanying SSSGs in their evolution.
This will allow to test directly small scale substructure formations
theories. High resolution spectroscopy will give the necessary
information (line--strength indices etc.) about single members
evolution.
 
\acknowledgments

We are deeply indebted to Dr. Luca Reduzzi which performed
observations. RR acknowledges the kind hospitality of the Institut
f\"ur Astronomie der Universit\"at Wien during the preparation of the
paper. LT and WWZ aknowledge the support of the Austrian Science Fund
(project P14783).  The research has made use of the NASA/IPAC
Extragalactic Database (NED) which is operated by the Jet Propulsion
Laboratory, California Institute of Technology, under contract with
National Aeronautics and Space Administration.



\clearpage


\begin{figure}
\caption{Bessel R band images of the SSSG sample. 
         Top row: SSSG~1 (A), SSSG~2 southern part (B), SSSG~2 northern part (C), SSSG~3 (D),
         central row: SSSG~4 (E), SSSG~5 (F), SSSG~6 southern part (G), SSSG~6 northern part (H)
         and bottom row: SSSG~7 (I), SSSG~8 (J), SSSG~9 (K). North is on the top, 
         East to the left. The field of view is 3.8\arcmin $\times$3.8\arcmin.
         \label{figure-1}}
\end{figure}

\begin{figure}
\caption{Spectral energy distribution for SSSG members: early--type galaxies (top row, left and
         central panel, late--type galaxies (top row, right panel and bottom row all panels)
         Main features are indicated. Spectra have been de-redshifted to the rest frame. 
         The galaxy spectra are displaced by a constant value of 
         5$\times$10$^{-14}$ erg~cm$^{-2}$~s$^{-1}$~{\AA}$^{-1}$, from each other for clarity.
         \label{figure-2}}
\end{figure}

\begin{figure}
\caption{Surface photometry of early--type SSSG members: SSSG~1a, SSSG~1b, SSSG~2a
         (top row), SSSG~2b; SSSG~3a, SSSG~3b (center row); SSSG~5a, SSSG~7a, SSSG~9b
         (bottom row). Each panel displays surface brightness (R mag~arcsec$^{-2}$), ellipticity
         (1-$b/a$), 
         position angle, and Fourier coefficients $a_4$ and $b_4$ as a function of radius
         \label{figure-3}}
\end{figure}

\begin{figure}
\caption{Surface photometry of late--type SSSG members: SSSG~2c, SSSG~2d, SSSG~5b
         (top row), SSSG~6a; SSSG~6b, SSSG~7b (center row); SSSG~8b, SSSG~9a
         (bottom row). Each panel displays surface brightness (R mag~arcsec$^{-2}$), 
         ellipticity (1-$b/a$), 
         position angle as a function of radius
         \label{figure-4}}
\end{figure}

\begin{figure}
\plotone{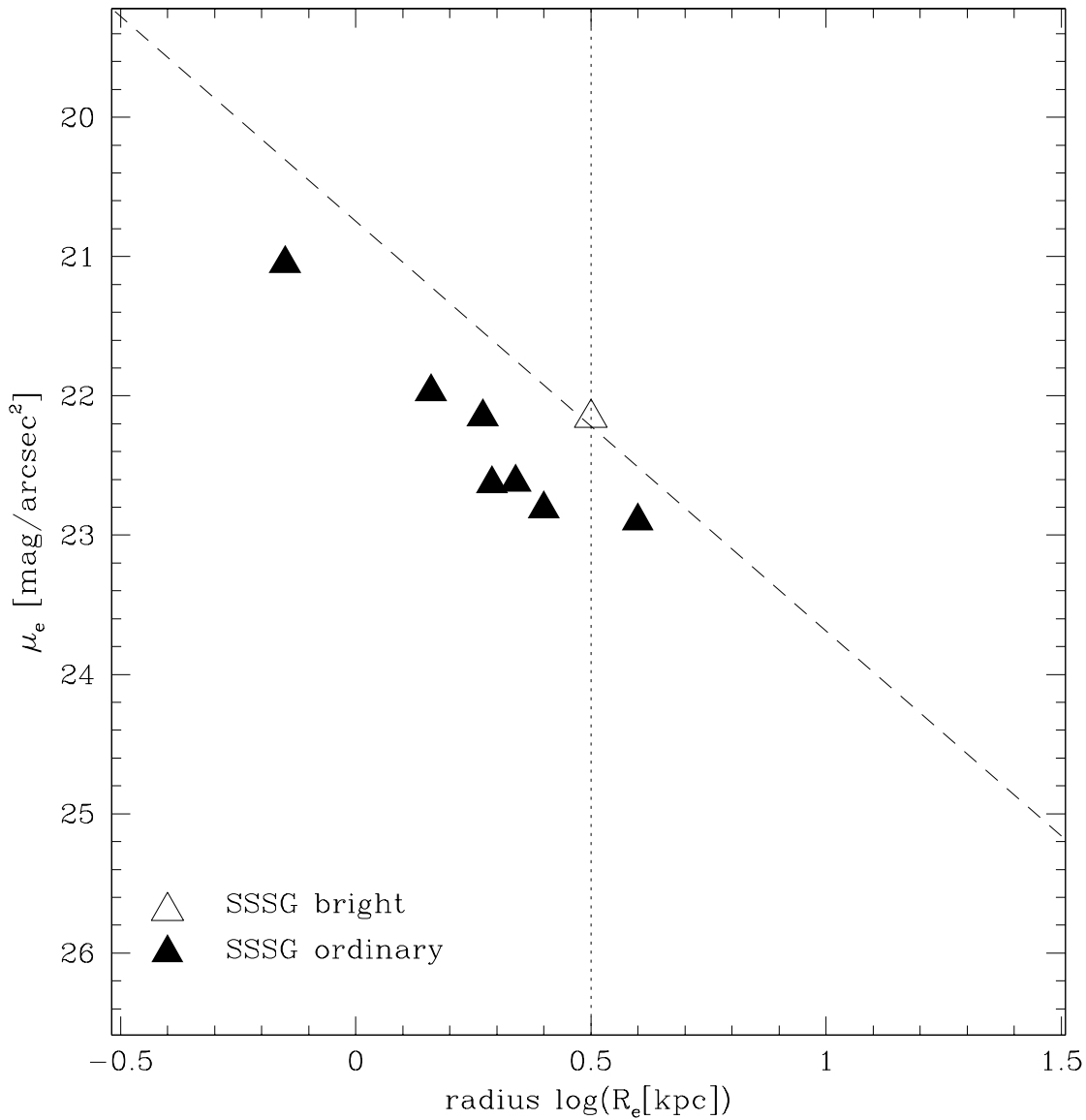}
\caption{Effective radius, in kpc, versus the effective B surface brightness. 
         The diagonal long-dashed line marks the HK87 relation, differentiating between
         bright and ordinary galaxies. The vertical dotted line at log R$_e$=0.5
         i.e. 3 kpc (H$_0$=70 km~s$^{-1}$~Mpc$^{-1}$) separates luminous from ordinary 
         ellipticals  according to \citet{cap92} \label{figure-5}}
\end{figure}

\clearpage 

\begin{figure}
\plottwo{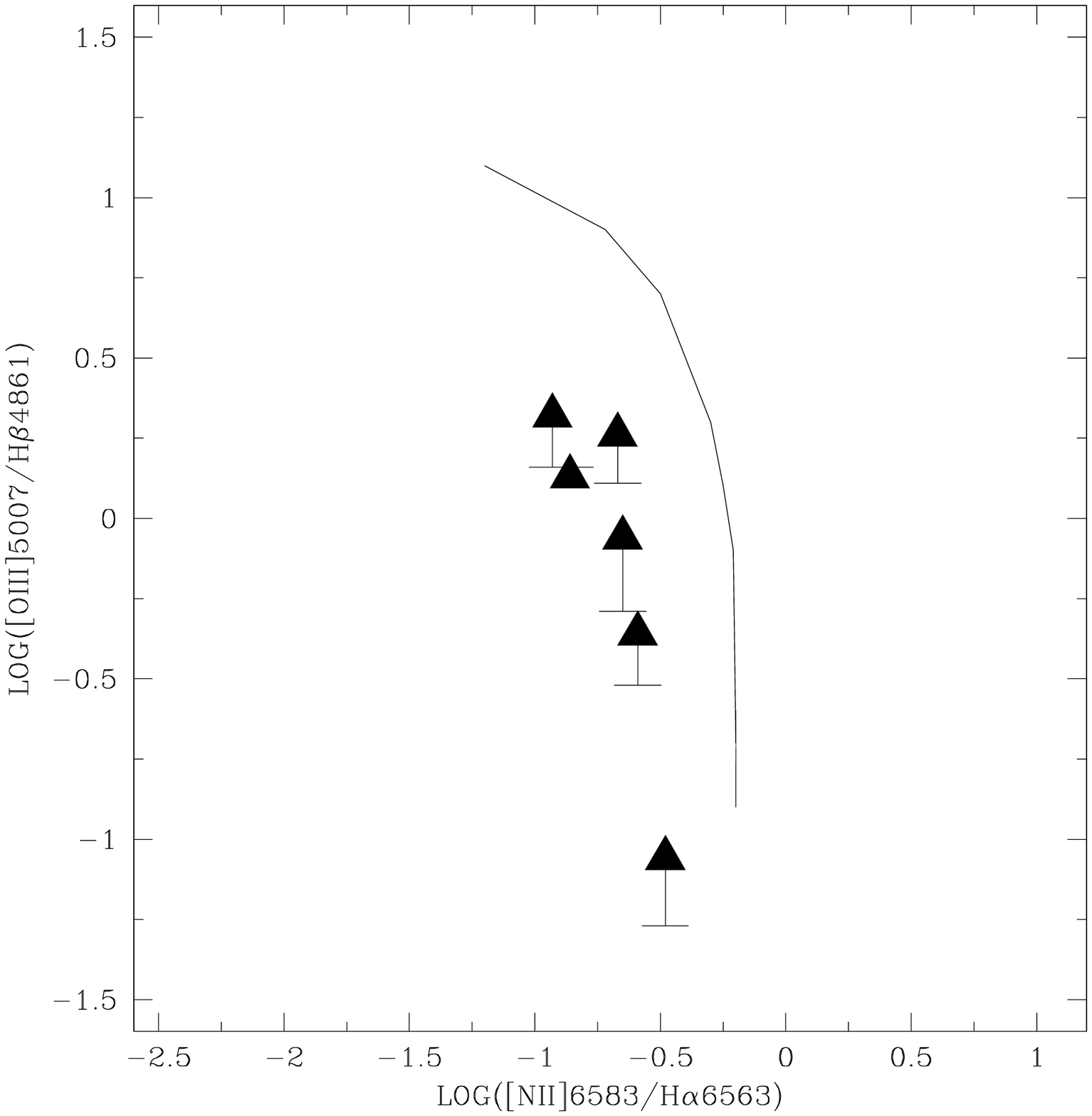}{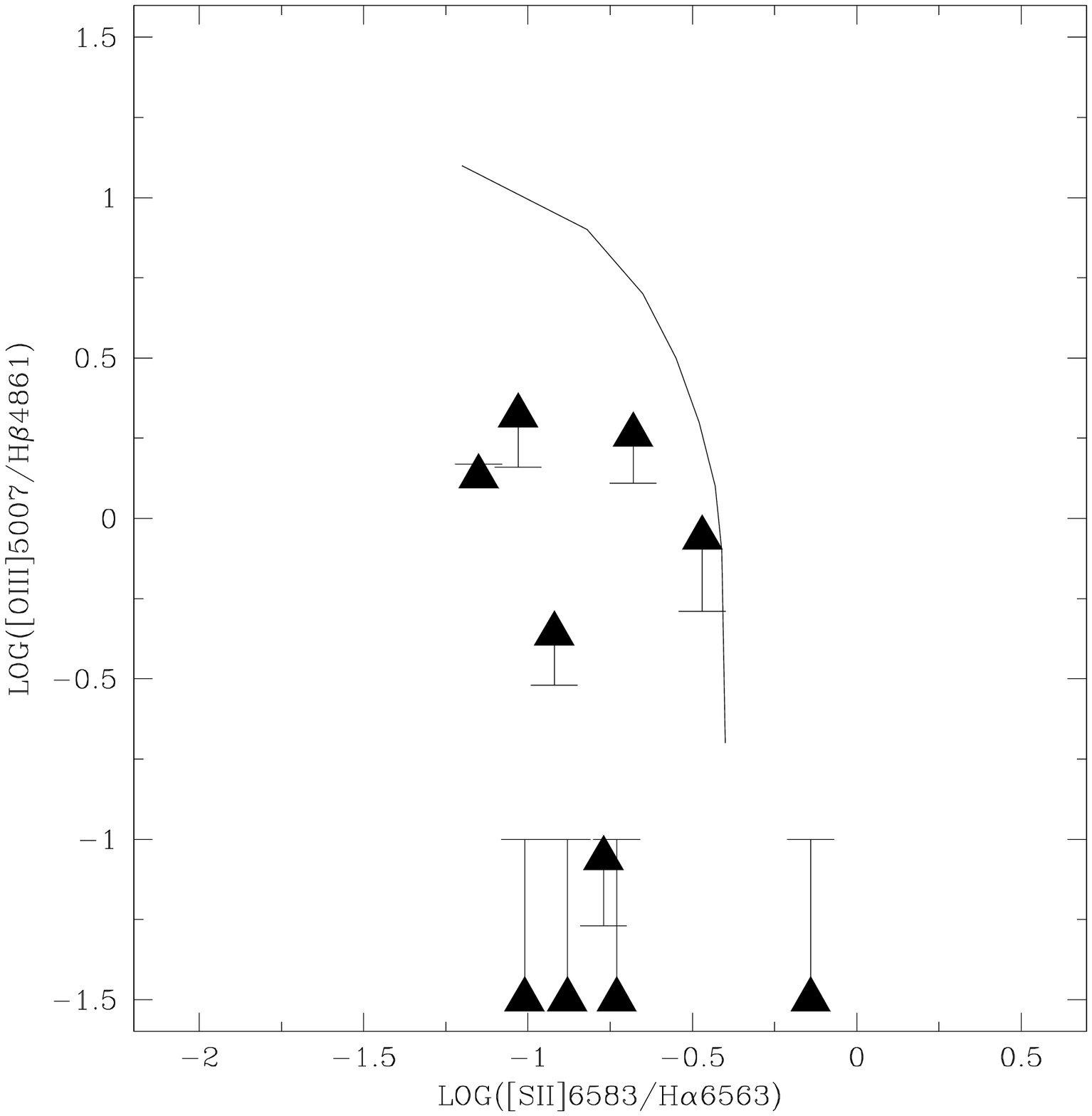}
\caption{[OIII]/H$\beta$ vs.[NII]/H$\alpha$ (left panel) and [OIII]/H$\beta$   
         vs. [SII]/H$\alpha$ (right panel) intensity ratios (solid curve divides 
         AGNs from HII region--like, same scale as has been used in Veilleux \& 
         Osterbrock, 1987). \label{figure-6}}
\end{figure}






\clearpage

\begin{deluxetable}{rccccrl}
\tabletypesize{\scriptsize}
\tablecaption{Parameters and environment of SSSG sample. \label{table-1}}
\tablewidth{0pt}
\tablehead{
  \colhead{SSSG}         & \colhead{V$_{hel}$}  & \colhead{n$_{CGA}$}
& \colhead{n$_{600kpc}$} & \colhead{n$_{1Mpc}$} 
& \colhead{$\sigma_v$}    & \colhead{Notes}\\
  \colhead{(1)} & \colhead{(2)} & \colhead{(3)}
& \colhead{(4)} & \colhead{(5)} & \colhead{(6)} & \colhead{(7)}\\
}
\startdata
 1  &  6588  & 4    & 1  & 5 & 172  & WBL 637 \\
 2  &  6688  & 4    & 2  & 3 & 113  & WBL 642 \\
 3  &  4578  & 4    & 0  & 5 & 168  &         \\
 4  &  5769  & 2    & 0  & 0 &   2  & KPG 551 \\
 5  &  8526  & 2    & 0  & 1 & 104  &         \\
 6  &  6297  & 3    & 0  & 1 & 132  &         \\
 7  &  8824  & 2    & 0  & 0 &  74  &         \\
 8  &  5433  & 2    & 1  & 3 &  88  &         \\
 9  &  4739  & 3    & 1  & 4 & 108  & \\
10  &  3630  & 2    & 0  & 0 &   9  & RR23\\
11  &  6081  & 2    & 1  & 4 & 489  & RR45 \\
\enddata
\tablenotetext{1}{\footnotesize SSSG identification}
\tablenotetext{2}{\footnotesize heliocentric velocity of the group [km s$^{-1}$]}
\tablenotetext{3}{\footnotesize number of group members within $r < 200$ h$_{100}^{-1}$~ kpc} 
\tablenotetext{4}{\footnotesize number of group members within $r < 600$ h$_{100}^{-1}$~ kpc}
\tablenotetext{5}{\footnotesize number of group members within $r < 1$ h$_{100}^{-1}$~ Mpc}
\tablenotetext{6}{\footnotesize velocity difference of the pair [km s$^{-1}$]}
\tablenotetext{7}{\footnotesize other group identifications}
\end{deluxetable}

\begin{deluxetable}{rrrlcl}
\tabletypesize{\tiny}
\tablecaption{Salient spectroscopic properties of the SSSG members.
              \label{table-2}}
\tablewidth{0pt}
\tablehead{
  \colhead{SSSG}   & \colhead{$\alpha$}  & \colhead{$\delta$}
& \colhead{other}  & \colhead{morphol.}  & \colhead{V$_{hel}$} \\
  \colhead{ident.} & \colhead{(2000)}    & \colhead{(2000)}
& \colhead{ident.} & \colhead{Type}      & \colhead{km~s$^{-1}$} \\
}
\startdata
 1a & 17 17 25.2 & 07 41 43  & CGCG 054-019   & S0                 & 6730$\pm$12         \\
  b & 17 17 33.4 & 07 39 43  & CGCG 054-020   & E                  & 6368$\pm$13         \\
    & 17 17 13.3 & 07 44 29  & CGCG 054-018   & \nodata            & 6536$^{4)}$         \\
    & 17 17 44.9 & 07 36 30  & UGC 10789      & \nodata            & 6719$^{4)}$         \\
 2a & 17 31 55.0 & 06 29 00  & CGCG 055-003   & E/S0               & 6846$\pm$13         \\
  b & 17 31 54.2 & 06 30 07  & CGCG 055-004   & E/S0               & 6687$\pm$15         \\
  c & 17 31 57.0 & 06 28 09  & CGCG 055-005   & SBc                & 6632$\pm$174$^{2)}$ \\
  d & 17 31 58.8 & 06 31 56  & CGCG 055-006   & Sc                 & 6587$\pm$34$^{2)}$  \\
 3a & 20 32 36.7 & 09 53 02  & NGC 6927A      & E/S0               & 4568$\pm$191$^{2)}$ \\
  b & 20 32 38.2 & 09 54 59  & NGC 6927       & S0:Sp$^{1)}$       & 4344$\pm$206$^{3)}$ \\
    & 20 32 50.2 & 09 55 38  & NGC 6928       & SB(s)ab$^{1)}$     & 4707 $^{4)}$        \\
    & 20 32 58.8 & 09 52 28  & NGC 6930       & SB(s)ab?$^{1)}$    & 4694$^{4)}$         \\
 4a & 20 59 46.9 &-01 53 16  & UGC 11657      & Irr                & 5768$\pm$45$^{2)}$  \\
  b & 20 59 48.3 &-01 52 23  & UGC 11658      & SAB(rs)pec?$^{1)}$ & 5770$\pm$7$^{2)}$   \\
 5a & 22 01 01.7 & 08 06 34  &A2158+0752      & S0                 & 8474$\pm$11         \\
  b & 22 01 10.8 & 08 07 32  &A2158+0753      & S                  & 8578$\pm$6$^{2)}$   \\
 6a & 00 02 34.8 &-03 42 38  &MCG -01-01-024  & SB(s)bc?$^{1)}$    & 6448$\pm$30$^{2)}$  \\
  b & 00 02 38.5 &-03 37 51  &MGC -01-01-025  & S                  & 6242$\pm$12$^{2)}$  \\
    & 00 02 48.7 &-03 36 21  &MCG -01-01-026  & S                  & 6202$^{4)}$         \\
 7a & 00 03 22.3 &-10 46 14  &MGC -02-01-012  & S0                 & 8861$\pm$11         \\
  b & 00 03 32.1 &-10 44 41  & NGC 7808       &(R')SA0$^{1)}$      & 8787$\pm$19         \\
 8a & 01 51 34.0 &-08 23 56  &MCG -02-05-065  & SB                 & 5389$\pm$2$^{2)}$   \\
  b & 01 51 27.0 &-08 30 20  & NGC 0707       & (R')SAB(s)0$^{1)}$ & 5477$\pm$2$^{2)}$   \\
 9a & 02 37 34.7 &-11 01 34  & NGC 1010       & SB                 & 4588$\pm$11         \\
  b & 02 37 38.9 &-11 00 20  & NGC 1011       & E0                 & 4754$\pm$17         \\
    & 02 37 49.8 &-11 00 39  & NGC 1017       & S                  & 4876$^{4)}$         \\
10a & 01 14 20.1 &-55 24 02  & NGC 0454 NED01 & Irr                & 3626$\pm$2$^{2)}$   \\
  b & 01 14 25.2 &-55 23 47  & NGC 0454 NED02 & Irr                & 3635$\pm$2$^{2)}$   \\
11a & 02 06 22.2 &-36 18 01  & ESO 354- G 036 & Sc$^{1)}$          & 6325$\pm$17$^{2)}$  \\
  b & 02 06 53.1 &-36 27 08  & NGC 824        & SB(RS)b$^{1)}$     & 5836$\pm$10$^{2)}$  \\ 
\enddata
\tablenotetext{1}{morphological type from NED}
\tablenotetext{2}{heliocentric velocity is determined from averaged gaussian fits to single emission lines}
\tablenotetext{3}{heliocentric velocity is determined from averaged gaussian fits to single absorption lines}
\tablenotetext{4}{heliocentric velocity from NED}
\end{deluxetable}

\begin{deluxetable}{rrcl|rrcl|rrcl}
\tabletypesize{\scriptsize}
\tablecaption{Spectroscopic observations \label{table-3}}
\tablewidth{0pt}
\tablehead{
  \colhead{SSSG}   & \colhead{PA} & \colhead{time} & \colhead{objects} 
& \colhead{SSSG}   & \colhead{PA} & \colhead{time} & \colhead{objects}   
& \colhead{SSSG}   & \colhead{PA} & \colhead{time} & \colhead{objects}  \\
  \colhead{(1)} & \colhead{(2)} & \colhead{(3)} & \colhead{(4)}  
& \colhead{(1)} & \colhead{(2)} & \colhead{(3)} & \colhead{(4)}  
& \colhead{(1)} & \colhead{(2)} & \colhead{(3)} & \colhead{(4)} \\ 
}
\startdata
{\bf 1} &  134 & 30 & a, b & {\bf 5}  &  67 & 30 & a, b & {\bf 9}  & 41 & 30 & a, b \\
{\bf 2} &  147 & 30 & a, c & {\bf 6}  &   4 & 30 & a    & {\bf 10} &    & 30 & a, b \\
{\bf 2} &   30 & 30 & b, d & {\bf 6}  &  98 & 30 & b    & {\bf 11} &    & 30 & a    \\
{\bf 3} &   11 & 30 & a, b & {\bf 7}  &  56 & 30 & b    &          &    &    &      \\
{\bf 4} &  22  & 30 & a, b & {\bf 8}  &  62 & 30 & b    &          &    &    &      \\ 
\enddata
\tablenotetext{1}{SSSG identification}
\tablenotetext{2}{position angle of the slit [deg]}
\tablenotetext{3}{exposure time [min]}
\tablenotetext{4}{SSSG member galaxies covered by the slit}
\end{deluxetable}

\begin{deluxetable}{rrrrrrrcccrr}
\tabletypesize{\scriptsize}
\tablecaption{Emission line fluxes and SFRs \label{table-4}}
\tablewidth{0pt}
\tablehead{
  \colhead{SSSG}      
& \colhead{H$\beta$} & \colhead{[OIII]N1}
& \colhead{[OIII]N2} & \colhead{H$\alpha$}
& \colhead{[NII]}    & \colhead{[SII]} 
& \colhead{log$\frac{[OIII]}{H\beta}$}
& \colhead{log$\frac{[NII]}{H\alpha}$}
& \colhead{log$\frac{[SII]}{H\alpha}$}
& \colhead{H$\alpha$/H$\beta$} 
& \colhead{SFR}\\  
  \colhead{ }      
& \colhead{(1)} & \colhead{(1)}
& \colhead{(1)} & \colhead{(1)}
& \colhead{(1)} & \colhead{(1)} 
& \colhead{ }
& \colhead{ }
& \colhead{ }
& \colhead{ } 
& \colhead{(2)}\\  
}
\startdata
 2c & \nodata$^3$ & \nodata & \nodata &   67.6 & 37.6 & \nodata~   & \nodata&  --0.26 & \nodata & \nodata&  8 \\
 2d & \nodata$^3$ & \nodata & \nodata &   25.6 &  8.8 &   7.5      & \nodata&  --0.47 & --0.73  & \nodata&  3 \\
 4a & 120.0       & 81.5    & 250.0   &  495.0 & 58.0 &  54.6      & 0.32   &  --0.93 & --1.03  & 4.13   & 43 \\
 4b & 85.0        & 17.5    &  37.0   &  348.0 & 89.4 &  41.4      & --0.36 &  --0.59 & --0.92  & 4.09   & 30 \\
 5b & \nodata~    & \nodata & \nodata &   25.9 & 17.1 & \nodata~   & \nodata&  --0.18 & \nodata & \nodata&  5 \\
 6a & \nodata$^3$ & \nodata & \nodata &   23.2 & 49.9 &  16.8      & \nodata&  0.33   & --0.14  & \nodata&  3 \\
 6b & 22.8        & 11.5    &  20.0   &  110.0 & 24.9 &  37.7      & --0.06 &  --0.65 & --0.47  & 4.82   & 11 \\
 7b & \nodata$^3$ & \nodata & \nodata &   14.8 & 16.5 & \nodata$^3$& \nodata&  0.05   & \nodata & \nodata&  3 \\
 8b & \nodata~    & \nodata &  36.1   &   23.4 & 21.0 &   3.1      & \nodata&  --0.05 & --0.88  & \nodata&  2 \\
 9a & 29.7        &  2.0    &   2.6   &  137.0 & 45.2 &  23.1      & --1.06 & --0.48  &  --0.77 &4.61    &  8 \\
 9b & \nodata$^3$ & \nodata & \nodata &   30.3 &  2.1 & \nodata~   & \nodata& --1.15  & \nodata & \nodata&  2 \\
10a & 575.0       &  272.0  & 769.0   & 1490.0 &204.0 & 105.0      & 0.13   & --0.86  &  --1.15 & 2.6    & 51 \\
10b & \nodata$^3$ & \nodata &  82.8   &  562.0 &566.0 &  54.6      & \nodata& 0.003   & --1.01  & \nodata& 19 \\
11a & 15.7        & 27.6    &  28.7   &   63.5 & 13.5 &  13.4      & 0.26   & --0.67  & --0.68  & 4.04    & 6 \\
\enddata
\tablenotetext{1}{Line fluxes are given in units of 10$^{-14}$ erg~cm$^{-2}$~s$^{-1}$}
\tablenotetext{2}{Star formation rate [M$_\odot$~yr$^{-1}$]}
\tablenotetext{3}{Spectral feature in absorption}
\end{deluxetable}

\begin{deluxetable}{rrrccccccl}
\tabletypesize{\scriptsize}
\tablecaption{Salient properties from the surface photometry of the SSSG members
              \label{table-5}}
\tablewidth{0pt}
\tablehead{
  \colhead{SSSG}    & \colhead{FWHM}   & \colhead{Exp. T.}   
& \colhead{R$_T$}   & \colhead{M$_R$}  & \colhead{r$_e$}
& \colhead{$\mu_e$} & \colhead{$\mu_0$} \\
  \colhead{(1)}     & \colhead{(2)} & \colhead{(3)}
& \colhead{(4)}     & \colhead{(5)}      
& \colhead{(6)}     & \colhead{(7)} & \colhead{(8)} \\
}
\startdata		
1a&1.4     & 30         & 13.13 & -20.89 & 12.30   & 22.02   & 18.73  \\
 b&        &            & 13.25 & -20.84 & 7.00    & 20.66   & 18.67  \\
2a&1.2/1.4 &2$\times$30 & 13.96 & -20.13 & 4.11    & 20.65   & 18.82  \\
 b&        &            & 14.21 & -19.92 & 3.08    & 20.47   & 18.60  \\
 c&        &            & 14.15 & -19.96 & \nodata & \nodata & 20.29  \\
 d&        &            & 14.30 & -19.79 & \nodata & \nodata & 20.70  \\
3a&1.4     & 30         & 15.02 & -18.21 & 2.31    & 19.55   & 19.02  \\
 b&        &            & 13.91 & -19.25 & 8.39    & 21.31   & 18.56  \\
4a&1.8     &3$\times$20 & 13.75 & -20.05 & \nodata & \nodata & 20.14  \\
 b&        &            & 13.07 & -20.74 & \nodata & \nodata & 20.21  \\
5a&1.2     & 30         & 14.83 & -19.78 & 3.74    & 21.12   & 19.30  \\
 b&        &            & 15.52 & -19.15 & \nodata & \nodata & 20.44  \\
7a&1.4     & 30         & 13.19 & -21.55 & 5.47    & 20.40   & 18.21  \\
 b&        &            & 12.54 & -22.17 & 12.30   & 21.89   & 18.49  \\
8a&1.4     & 40         & 14.14 & -21.00 & \nodata & \nodata & 19.72  \\
 b&        &            & 13.73 & -19.83 & \nodata & \nodata & 19.61  \\
9a&1.4     & 30         & 12.82 & -20.54 & 12.29   & 21.40   & 19.67  \\
 b&        &            & 13.35 & -20.01 & 6.02    & 21.00   & 18.69  \\
\enddata
\tablenotetext{1}{SSSG and group member galaxy identification}
\tablenotetext{2}{seeing as measured FWHM of stellar images in the frame[$''$] }
\tablenotetext{3}{exposure time [min]}
\tablenotetext{4}{apparent total magnitude $m_R$}
\tablenotetext{5}{absolute total magnitude $M_R$}
\tablenotetext{6}{effective radius [$''$]}
\tablenotetext{7}{surface brightness $\mu_e$(R) measured at the effective radius [mag~arcsec$^{-2}$]}
\tablenotetext{8}{central surface brightness $\mu_0$(R) [mag~arcsec$^{-2}$]}
\end{deluxetable}

\begin{deluxetable}{lcc}
\tabletypesize{\scriptsize}
\tablecaption{Arm classes of the spiral members. \label{table-6}}
\tablewidth{0pt}
\tablehead{
\colhead{SSSG \#} & \colhead{AC type} &  \colhead{Notes} \\
}
\startdata
2c &   9 & bar    \\
2d &   3 &        \\
5b &  12 &        \\
7b &   8 &        \\
8b &  12 & bar    \\
9a &   6 & bar    \\
\enddata
\end{deluxetable}


\begin{thebibliography}{}

\bibitem[Barnes(1996)]{bar96} 
        Barnes, J. 1996, 
        Galaxies: Interactions and Induced Star Formation,  
        Saas--Fee Advanced Course 26, 275  
\bibitem[Combes et al.(1994)]{com94} 
        Combes, F., Prugniel, P., Rampazzo  R. \& Sulentic, J. W. 
        1994, \aap, 281, 725
\bibitem[Capaccioli et al.(1992)]{cap92} 
        Capaccioli, M., Caon, N., D'Onofrio, M. 1992, 
        Structure, Dynamics and Chemical Evolution of Early--Type Galaxies, 
        ESO-EIPC Workshop, Danziger J. et al., 43 (CCD92)
\bibitem[Coziol et al.(2000)]{coz00} 
        Coziol, R., Iovino, A. \& de Carvalho, R. R. 2000, \aj, 120, 47
\bibitem[Diaferio et al.(1994)]{dia94} 
        Diaferio, A. Geller, M.J. \& Ramella, M. 1994, \aj, 107, 868
\bibitem[Elmegreen \& Elmegreen(1982)]{elm82} 
        Elmegreen, D. M. \& Elmegreen, B. G. 1982, \mnras, 201, 1021
\bibitem[Elmegreen \& Elmegreen(1987)]{elm87} 
        Elmegreen, D. M. \& Elmegreen, B. G. 1987, \apj, 314, 3
\bibitem[Focardi \& Kelm(2002)]{foc02} 
        Focardi, P. \& Kelm, B. 2002, \aap, 391, 35
\bibitem[Forman \& Jones(1982)]{for82} 
        Forman, W. \& Jones, C. 1982, \araa, 20, 547
\bibitem[Governato, Tozzi \& Cavaliere(1996)]{gov96} 
        Governato, F., Tozzi, P. \& Cavaliere, A. 1996, \apj, 458, 18
\bibitem[Haynes, Giovanelli \& Chincarini(1984)]{hay84} 
        Haynes, M.P., Giovanelli, R. \& Chincarini, G. 1984, \araa, 22, 445
\bibitem[Hamabe \& Kormendy(1987)]{ham87} 
        Hamabe, M. \&  Kormendy, J. 1987, 
        Structure and Dynamics of elliptical Galaxies, 
        IAU Symp. No. 127 (Princeton), ed. T. de Zeeuw, Dordrecht: Reidel, 
        379: (HK87)
\bibitem[Henricksen \& Cousineau(1999)]{hen99} 
        Henricksen M. \& Cousineau, S. 1999, \apj, 511, 595
\bibitem[Hickson(1997)]{hic97}
        Hickson P. 1997, \araa, 35, 357 
\bibitem[Huchra et al. (1992)]{huc92}
        Huchra, J.P, Geller, M.J., Clemens C.M., Tokarz S.P. \& Michel A. 
        1992, Bull. Inf. CDS, 41,31
\bibitem[Jedrzejewski(1987)]{jed87} 
        Jedrzejewski, R.  1987, MNRAS 226, 747
\bibitem[Junqueira, de Mello, \& Infante(1998)]{jun98} 
        Junqueira, S., de Mello, D. F. \& Infante, L., 1998, \aaps, 129, 69
\bibitem[Karachentsev(1972)]{kar72}
        Karachentsev, I. D. 1972, 
        Catalogue of isolated pair of galaxies in the northern hemisphere, 
        Soob-shch. Spets. Astrofiz. Obs. 7,3
\bibitem[Keel(1996)]{kee96} 
        Keel, W. C. 1996, \aj, 111, 696
\bibitem[Kennicutt(1996)]{ken96} 
        Kennicutt, R. C. 1996, 
        Galaxies: Interactions and Induced Star Formation,  
        Saas--Fee Advanced Course 26, 1
\bibitem[Kennicutt(1998)]{ken98} 
        Kennicutt, R.C. 1998, \araa, 36, 189
\bibitem[Kelm, Focardi \& Palumbo(1998)]{kel98}  
        Kelm, B., Focardi, P. \& Palumbo, G.G.C. 1998, \aap,  335, 912.
\bibitem[Kelm, Focardi \& Zampieri(2003)]{kel03}
        Kelm, B., Focardi P. \& Zampieri, A. 2003, 
        Galaxy Evolution III: From simple Approaches to self consistent Models, 
        ed. G. Hensler, Kluwer Academic Publisher, in press 
\bibitem[Laurikainen \& Salo(1995)]{lau95} 
        Laurikainen, E. \& Salo, H. 1995, \aap 293, 683
\bibitem[Longhetti et al.(1998a)]{lon98a} 
        Longhetti,  M., Rampazzo, R., Bressan, A. \& Chiosi, C. 1998a, \aap, 130, 251
\bibitem[Longhetti et al.(1998b)]{lon98b} 
        Longhetti,  M., Rampazzo, R., Bressan, A. \& Chiosi, C. 1998b, \aap, 130, 267
\bibitem[Longhetti et al.(1999)]{lon99} 
        Longhetti, M., Bressan, A., Chiosi, C. \& Rampazzo, R. 1999, \aap, 345, 519
\bibitem[Longhetti et al.(2000)]{lon00} 
        Longhetti, M., Bressan, A., Chiosi, C. \&  Rampazzo, R. 2000, \aap, 353, 917
\bibitem[Monaco et al.(1994)]{mon94} 
        Monaco, P., Giuricin, G., Mardirossian, F., Mezzetti, M. et al. 1994, \apj, 436, 576
\bibitem[Moore et al.(1996)]{moo96} 
        Moore, B., Katz, N., Lake, G., Dressler, A. \& Oemler, A., Jr. 1996, \nat, 379, 613
\bibitem[Mulchaey \& Zabludoff(1999)]{mul99} 
        Mulchaey, J. S. \& Zabludoff, A. I. 1999, \apj, 514, 33
\bibitem[Mulchaey(2000)]{mul00} 
        Mulchaey, J. S. 2000, \araa, 38, 289
\bibitem[Navarro(1990)]{nav90} 
        Navarro, J. F. 1990, \mnras, 242, 311
\bibitem[Noguchi \& Ishibashi(1986)]{nog86} 
        Noguchi, M. \& Ishibashi, S. 1986, \mnras, 219, 305
\bibitem[Pierfederici, Rampazzo \& Reduzzi(2000)]{pie00} 
        Pierfederici, F., Rampazzo, R. \& Reduzzi, L. 2000, \aplett, Vol. 40, 84.
\bibitem[Ponman et al.(1996)]{pon96} 
        Ponman, T. J., Bourner, P. D. J., Ebeling, H. \& B\"ohringer, H. 1996, \mnras, 283, 690
\bibitem[Rafanelli, Violato \& Baruffolo(1995)]{raf95} 
        Rafanelli, P., Violato, M. \& Baruffolo, A. 1995, \aj, 109, 1546
\bibitem[Rampazzo \& Sulentic(1992)]{ram92} 
        Rampazzo, R. \& Sulentic, J. W. 1992, \aap 259, 43
\bibitem[Rampazzo et al.(2000)]{ram00} 
        Rampazzo, R., D'Onofrio, M., Bonfanti, P., Longhetti, M. \& Reduzzi, L. 2000, 
        \aplett, Vol. 40, 63
\bibitem[Reduzzi \& Rampazzo(1995)]{red95} 
        Reduzzi, L. \& Rampazzo, R. 1995, \aplett, Vol. 30., 1: RR95
\bibitem[Reduzzi, Longhetti \& Rampazzo(1996)]{red96} 
        Reduzzi, L., Longhetti, M. \& Rampazzo,  R. 1996, \mnras, 282, 149
\bibitem[Schweizer(1992)]{sch92} 
        Schweizer, F. 1992, 
        Structure, Dynamics and Chemical Evolution of Early--type  Galaxies, 
        ESO--EIPC Workshop, Danziger et al., 651
\bibitem[Schweizer(1996)]{sch96} 
        Schweizer F. 1996 
        Galaxies: Interactions and Induced Star Formation,  
        Saas--Fee Advanced Course 26, 105
\bibitem[Struble \& Rood(1999)]{str99}
        Struble, M. F. \& Rood, H. J. 1999, \apjs, 125, 35
\bibitem[Sulentic \& de Mello Rabaca(1993)]{sul93}
        Sulentic J.W. \& de Mello Rabaca, D. 1993, \apj, 410 520 
\bibitem[Thomson \& Wright(1990)]{tho90} 
        Thomson, R. C. \& Wright, A. E. 1990, \mnras, 247, 122
\bibitem[Thomson(1991)]{tho91} 
        Thomson, R. C.  1991, \mnras, 253, 256
\bibitem[Trinchieri \& Rampazzo(2001)]{tri01} 
	Trinchieri, G. \& Rampazzo, R. 2001, \aap, 374, 454
\bibitem[Veilleux \& Osterbrock(1987)]{vei87} 
        Veilleux, S. \& Osterbrock D. E. 1987, \apjs, 63, 295
\bibitem[Verdes--Montenegro et al.(1998)]{ver98} 
        Verdes--Montenegro, L.,  Yun, M. S., Perea, J., del Olmo, A. \& Ho, P. T. P.  
        1998, \apj, 497, 89
\bibitem[Weil \& Hernquist(1993)]{wei93}
        Weil, M.L. \& Hernquist, L. 1993, \apj, 405, 142
\bibitem[Xu \& Sulentic(1991)]{xus91} 
        Xu, C. \& Sulentic J. W. 1991, \apj, 374, 407 
\bibitem[White et al.(1999)]{whi99}
        White, R. A., Blinton, M., Bhavsar, S. P. et al. 1999, \aj, 118, 2014
\bibitem[Zabludoff \& Mulchaey(1998)]{zab98} 
        Zabludoff, A. \& Mulchaey, J. 1998, \apj, 498, L5
\bibitem[Zepf \&  Whitmore(1991)]{zep91} 
        Zepf, S. E. \&  Whitmore, B. C.  1991, \apj, 383, 542
\bibitem[Zepf(1993)]{zep93} 
        Zepf, S. E. 1993, \apj, 407, 448

\end{thebibliography}
\end{document}